\documentclass[article]{siamart250211}
\usepackage[caption=false]{subfig}
\usepackage{xcolor}
\title{Comparison of Extended Lubrication Theories for Stokes Flow}

\author{Sarah Dennis\thanks{Department of Mathematics, Brandeis University, Waltham MA (\email{sarahdennis@brandeis.edu})} \and Thomas G. Fai\thanks{Department of Mathematics and Volen Center for Complex Systems, Brandeis University, Waltham MA (\email{tfai@brandeis.edu})}}

\begin{document}
\maketitle
\begin{abstract}
Lubrication theory makes use of the assumptions of a long and thin fluid domain and a small scaled Reynolds number to formulate a linearized approximation to the Navier-Stokes equations. Extended lubrication theory aims to improve the model accuracy by relaxing these assumptions and including additional terms in the formulation. However, such models are sensitive to large surface gradients which lead the assumptions of the model to break down. In this paper, we present a formulation of extended lubrication theory, and compare our model with several existing models, along with the numerical solution to the Stokes equations. The error in pressure and velocity is characterized for a variety of fluid domain geometries. Our results indicate that the new solution is suitable for a wide range of geometries. The magnitude of surface variation and the length scale ratio are both important factors influencing the accuracy of the extended lubrication theory models.
\end{abstract}
\section{Introduction}
The classical lubrication assumptions of a long and thin fluid domain and a small scaled Reynolds number are used to justify neglecting the inertial velocity and cross-film pressure gradient terms in the Navier-Stokes equations. The resulting Reynolds equation is a dramatic simplification of the original governing equations. More recently, variations on the thin film assumption have given rise to higher-order approximations for lubrication theory \cite{venner_stokes_2003,marusic-paloka_second_2012,tavakol_extended_2017,takeuchi_extended_2019,sialmas_newtonian_2024,okazaki_fluid_2024}. These extended lubrication theory models retain certain simplifying assumptions as in classical lubrication theory, while nevertheless being applicable to a broader scope of geometries in the low Reynolds number limit.

Previous research on the flows modeled by lubrication theory has highlighted the sensitivity of lubrication theory to variations of the fluid film height. In particular, in the limit of zero Reynolds number flows, discrepancies between lubrication theory and Stokes flow increase with both the magnitude and frequency of large gradients in the height \cite{brown_applicability_1995,biswas_backward-facing_2004,dobrica_about_2009}. The pressure drop modeled by the Reynolds equation may be underestimated due to additional pressure losses at points of sudden expansion that are not accurately modeled \cite{dobrica_reynolds_2005,dobrica_about_2009}. Furthermore, notable flow features of corner recirculation seen in the Stokes solution are not accurately captured with the Reynolds equation \cite{armaly_experimental_1983,biswas_backward-facing_2004,shyu_numerical_2018}. {\color{black} Note that the references cited above should not be interpreted as an exhaustive review; perturbation approaches are a rich area of research in the field of fluid dynamics. The references \cite{venner_stokes_2003,marusic-paloka_second_2012,tavakol_extended_2017,takeuchi_extended_2019,sialmas_newtonian_2024,okazaki_fluid_2024} focus on Newtonian fluids and perturbation of the length scale ratio which quantifies the thin film assumption of classical lubrication theory. For further discussion of perturbation theories, including for non-Newtonian fluids, we direct interested readers to \cite{tshehla_flow_2013,housiadas_high-order_2016,synder_comparison_2018,ahmed_new_2021,boyko_pressure-driven_2022,abbaspur_analytical_2023,housiadas_pressure-drop_2024,okazaki_fluid_2024,ahmed_viscoelastic_2025,sialmas_general_2025} and references therein.}

In this paper, we consider classical lubrication theory (the Reynolds equation), and more recent variations on lubrication theory which we refer to as extended lubrication theory \cite{takeuchi_extended_2019,okazaki_fluid_2024} or perturbed lubrication theory \cite{venner_stokes_2003,marusic-paloka_second_2012,tavakol_extended_2017,sialmas_newtonian_2024} in the case of models based on perturbation theory. We assess the performance of these models at low Reynolds numbers by comparison to the Stokes equations. We propose a formulation of extended lubrication theory that builds on the model of Takeuchi \& Gu (2019) \cite{takeuchi_extended_2019}; our approach modifies the treatment of the velocity boundary conditions and ensures incompressibility. The pressure and velocity solutions from each model are compared in a variety of geometries: a logistic step, a backward facing step, and two versions of a triangular slider that we refer to as positive and negative based on the texture to be defined later on. Through these examples, we assess the sensitivity of each model to varying magnitudes of surface variation and discontinuities in the surface gradient in the low Reynolds number limit.

Our results indicate that the extended and perturbed lubrication theory models provide significant improvements on classical lubrication theory in the case of small to moderate surface variation. For the logistic step, our proposed solution in extended lubrication theory has the smallest error in velocity for small slopes, and the smallest error in pressure at moderate slopes. For the triangular slider, the perturbed solutions have significantly reduced error in pressure, particularly in the case of negative texturing. In general, we find that for geometries with a large surface gradient, the extended and perturbed solutions may over-correct the cross-film pressure gradient.

Source code is available at \href{https://github.com/sarah-dennis/extended-lubrication-theory}{github.com/sarah-dennis/extended-lubrication-theory}.

\section{Background}
\subsection{Lubrication Theory}
We begin with a review of classical lubrication theory. This treatment may be found in various classical textbooks \cite{leal_advanced_2007}, and we include it here to establish notation and to make the presentation self-contained.
The Navier-Stokes equations,
\begin{align}\label{n-s_x}
 \frac{\partial p}{\partial x} &= \eta \Big(\frac{\partial^2 u}{\partial x^2} + \frac{\partial^2 u}{\partial y^2}\Big) - \rho\Big( \frac{\partial u}{\partial t}+ u\frac{\partial u}{\partial x} + v \frac{\partial u}{\partial y}\Big),\\
 \label{n-s_y} \frac{\partial p}{\partial y} &= \eta \Big(\frac{\partial^2 v}{\partial x^2} + \frac{\partial^2 v}{\partial y^2}\Big) - \rho\Big(\frac{\partial v}{\partial t}+ u\frac{\partial v}{\partial x} + v \frac{\partial v}{\partial y} \Big),
\end{align}
subject to incompressibility,
\begin{equation}\label{incompressibility}
 \frac{\partial u}{\partial x} + \frac{\partial v}{\partial y} = 0,\end{equation}
serve as the governing equations for a two-dimensional incompressible fluid. The pressure is $p$, the velocity is $(u,v)$, and the kinematic viscosity is $\nu=\eta/\rho$, where $\eta$ is the bulk viscosity and $\rho$ is the constant density.

Lubrication theory provides a linearized approximation to the Navier-Stokes equations valid for geometries in which the fluid is constrained to a relatively thin domain. For the two-dimensional fluid domain $(x,y) \in [0,L_x]\times[0,h(x)]$, the characteristic length scales are $L_x>0$ and $L_y = \max h(x) > 0$. Given a prescribed constant flux $\mathcal{Q}\ne 0$, the characteristic velocities are $U_* = \mathcal{Q}/L_y$ and $V_* = \mathcal{Q}/L_x$, the characteristic time is $T_* = L_x/U_*$, and the characteristic pressure is $P_*=\eta U_*L_x/L_y^2$. The dimensionless variables are $X=x/L_x$, $Y=y/L_y$,  $H=h/L_y$, $U=u/U_*$, $V=v/V_*$, $T = t/T_*$ and $P=p/P_*$. The Reynolds number is $\text{Re}=\rho U_*L_x/\eta$ and the length scale ratio is $\varepsilon = L_y/L_x$. The Navier-Stokes equations \cref{n-s_x,n-s_y} in dimensionless terms are,
\begin{align}\label{n-s_x_dimless_re}
\frac{\partial P}{\partial X} &= \Big(\varepsilon^2\frac{\partial^2 U}{\partial X^2} + \frac{\partial^2 U}{\partial Y^2}\Big) - \varepsilon^2\text{Re}\Big( \frac{\partial U}{\partial T}+U\frac{\partial U}{\partial X} + V \frac{\partial U}{\partial Y} \Big),\\
 \label{n-s_y_dimless_re} \frac{\partial P}{\partial Y} &= \varepsilon^2 \Big(\varepsilon^2\frac{\partial^2 V}{\partial X^2} + \frac{\partial^2 V}{\partial Y^2}\Big) - \varepsilon^4 \text{Re}\Big( \frac{\partial V}{\partial T}+U\frac{\partial V}{\partial X} +V\frac{\partial V}{\partial Y}\Big),
\end{align}
with incompressibility \cref{incompressibility} which is unchanged.

We assume the no-slip boundary condition at the fluid-surface interfaces $Y=0$ and $Y=H(X)$. Without loss of generality, the velocity boundary conditions are,
\begin{align}
\label{reyn_bc_u} && U (X,0) = U_b && U(X,H) = 0, \\
\label{reyn_bc_v} && V (X,0) = 0&& V(X, H) = 0,
\end{align}
where $U_b=\mathcal{U}/U_*$ is the constant relative velocity between the surfaces. We prescribe the flux $\mathcal{Q}$ so that, in dimensionless terms, the velocity satisfies, \begin{equation}\label{flux}\int_0^{H(X)}U(X,Y)\,dY=1.\end{equation}
We also prescribe the fixed outlet pressure $P(1,0) = 0$. {\color{black}Because the flux is prescribed, the pressure drop $\Delta P$ is given from the solution. Conversely, one may prescribe the pressure drop, in which case the flux would be determined by the solution. We define the dimensionless pressure drop,   
\begin{equation}\label{dP}
    \Delta P = \int_0^{H(0)} P(0,Y)\,dY - \int_0^{H(1)} P(1,Y)\,dY,
\end{equation}
and the dimensional pressure drop,
\begin{equation}\label{dp}
    \Delta p = \frac{1}{L_y}\int_0^{h(0)} p(0,y)\,dy - \frac{1}{L_y}\int_0^{h(L_x)} p(L_x,y)\,dy,
\end{equation}
such that $\Delta P= \Delta p/P_*$.}

The lubrication assumptions $\varepsilon \ll 1$ and $\varepsilon^2 \text{Re}\ll 1$ characterize a long and thin fluid with a small scaled Reynolds number. These assumptions yield an approximation to \cref{n-s_x_dimless_re,n-s_y_dimless_re} with the resulting equations,
\begin{align}
\label{gap_x} \frac{\partial P_0}{\partial X}&=\frac{\partial^2 U_0}{\partial Y^2}, \\ 
\label{gap_y} \frac{\partial P_0}{\partial Y}&=0,
\end{align}
where $P_0$, $U_0$, $V_0$ denote the order $\varepsilon^0$ lubrication approximation.
Together with incompressibility \cref{incompressibility}, \cref{gap_x,gap_y} constitute the governing equations for lubrication theory. 

The velocity $U_0$ is determined through integration of \cref{gap_x} and applying the boundary conditions \cref{reyn_bc_u},
\begin{equation}
  \label{reyn_u} U_0(X,Y)=\frac{1}{2}\frac{d P_0}{d X} \Big(Y^2-H Y\Big) + U_b\frac{ \big(H-Y\big)}{H}.
 \end{equation}
The velocity $V_0$ is determined from \cref{reyn_u} and incompressibility \cref{incompressibility} with the boundary conditions \cref{reyn_bc_v},
\begin{equation}\label{reyn_v}
V_0(X,Y) =\frac{-1}{6}\frac{d^2 P_0}{d X^2}Y^3 + \frac{1}{2}\Bigg(\frac{1}{2}\Big(\frac{d^2 P_0}{d X^2}H + \frac{d P_0}{d X}\frac{dH}{d X}\Big) -\frac{U_b}{H^2}\frac{dH}{dX}\Bigg)Y^2.
\end{equation}
The condition of constant flux and the boundary condition $V_0(X,H)=0$ are satisfied exactly when $P_0$ satisfies the Reynolds equation,
\begin{equation}\label{reynolds}
     \frac{d}{dX}\Bigg[H^3\frac{dP_0}{dX}\Bigg] = 6U_{b}\frac{dH}{dX}.
\end{equation}

To solve the Reynolds equation, we assign the inlet pressure gradient according to the prescribed flux, along with the fixed outlet pressure,
\begin{align}\label{reyn_bc_p}&& \frac{dP_0}{dX}\bigg|_{0} = -12\big[H(0)\big]^{-3} +6U_b\big[H(0)\big]^{-2}, && P_0(1) =0. &&\end{align}
{\color{black} The inlet pressure gradient \cref{reyn_bc_p} is obtained through integration of $U_0$ from \cref{reyn_u} according to the flux condition \cref{flux}, evaluated at $X = 0$.}
Note that the flux appears in the scaling of the boundary velocity, $U_b=\mathcal{U}L_y/\mathcal{Q}$. Since the flux is prescribed, the inlet pressure $P_0(0)$ and the pressure drop $\Delta P_0$ are determined by the solution. The solution $P_0$ to the Reynolds equation is determined using a second-order accurate finite difference scheme, see \cref{app_reyn_fd}. 

\subsection{Perturbed Lubrication Theory}\label{sec_pert_re}
The dimensionless Navier-Stokes equations \cref{n-s_x_dimless_re,n-s_y_dimless_re} motivate the approach of an $\varepsilon$-perturbed solution \cite{venner_stokes_2003,marusic-paloka_second_2012,tavakol_extended_2017,sialmas_newtonian_2024}. Given a sufficiently small Reynolds number such that $\varepsilon^2 \text{Re}\ll 1$, perturbed lubrication theory assumes a solution of pressure and velocity of the form,
\begin{align}\label{pert_p}
    P(X,Y)&=P_{0} + \varepsilon^2 P_{2} + \varepsilon^4 P_{4},\\\label{pert_u}
   U(X,Y)&=U_{0} + \varepsilon^2 U_{2} + \varepsilon^4 U_{4},\\\label{pert_v}
    V(X,Y)&=V_{0} + \varepsilon^2 V_{2} + \varepsilon^4 V_{4}.
\end{align}
The order $\varepsilon^k$ solution ($\varepsilon^k$-PLT) assumes a long and thin domain characterized by $\varepsilon^k \sim \mathcal{O}(1)$ and such that higher powers of $\varepsilon$ are negligible. The $\varepsilon^0$-PLT solution $U_0$, $V_0$, $P_0$ is the solution of lubrication theory and the Reynolds equation.

On substituting the ansatz \cref{pert_p,pert_u,pert_v} into \cref{n-s_x_dimless_re,n-s_y_dimless_re} at order $\varepsilon^2$, the velocity and pressure satisfy the following reduced equations,
\begin{align}\label{pert_O2_x}
    \frac{\partial P_2}{\partial X}&=\frac{\partial^2 U_{0}}{\partial X^2}+\frac{\partial^2 U_{2}}{\partial Y^2}, \\ \label{pert_O2_y}
    \frac{\partial P_2}{\partial Y}&=\frac{\partial^2 V_{0}}{\partial Y^2}.
\end{align} Likewise at order $\varepsilon^4$,
\begin{align}\label{pert_O4_x}
    \frac{\partial P_4}{\partial X}&=\frac{\partial^2 U_{2}}{\partial X^2}+\frac{\partial^2 U_{4}}{\partial Y^2}, \\ \label{pert_O4_y}
    \frac{\partial P_4}{\partial Y}&=\frac{\partial^2 V_{2}}{\partial Y^2} + \frac{\partial^2 V_{0}}{\partial X^2}.
\end{align}
The velocity fields $(U_k, V_k)$ additionally all satisfy incompressibility \cref{incompressibility}.

The order $\varepsilon^0$ Reynolds solution $U_0$, $V_0$, and $P_0$ satisfies the boundary conditions \cref{reyn_bc_u,reyn_bc_v,reyn_bc_p} respectively. The higher-order velocities $U$ and $V$ also satisfy \cref{reyn_bc_u,reyn_bc_v}. The higher-order pressures $P$ satisfy $P(1,0)=0$, but do not necessarily satisfy the pressure gradient condition in \cref{reyn_bc_p}. Instead, we require that the flux $\mathcal{Q}$ be the same as the flux of the order $\varepsilon^0$ solution, such that, in dimensionless terms, $\int_0^H U_{0}\,dY = 1$ and  $\int_0^H U_{k>0}\,dY = 0$. Because the flux is prescribed, the average pressure drop $\Delta P_{\varepsilon^k\text{-PLT}}$ is determined by the solution. In general, the average pressure drops $\Delta P_{\varepsilon^2\text{-PLT}}=\Delta P_0 + \varepsilon^2 \Delta P_2$ and $\Delta P_{\varepsilon^4\text{-PLT}}=\Delta P_0 + \varepsilon^2 \Delta P_2 + \varepsilon^4 \Delta P_4$ will deviate from $\Delta P_0$. We determine $\Delta P_{k>0}$ through numerical integration according to \cref{dP}.

The method of perturbed solutions is appealing in that we obtain exact expressions for the velocities $U_k$ and $V_k$ involving only $Y$, the height function $H$ and its derivatives. The expressions for pressure $P_k$ do involve integrals requiring numerical integration for an arbitrary height function. The complete expressions for $U_k$, $V_k$ and $P_k$ at orders $\varepsilon^2$ and $\varepsilon^4$ are presented in the appendix \cref{app_pert}. Similar expressions up to various orders appear in \cite{venner_stokes_2003,
marusic-paloka_second_2012,tavakol_extended_2017}. Here we have provided corrected expressions for $V_2$ and $U_4$ in \cite{tavakol_extended_2017}, and we include an expression for $V_4$ which is not otherwise explicitly written in \cite{venner_stokes_2003,
marusic-paloka_second_2012,tavakol_extended_2017}. 

The method of perturbed solutions is also well suited to techniques such as diagonal Pad\'e approximants or the Shanks' transform \cite{andrianov_pade_2021,housiadas_improved_2023,sialmas_newtonian_2024,sialmas_general_2025}. These techniques may be used to accelerate and extend the domain of the convergence for the series solution. Although such numerical methods are out of the scope of the present work, we consider this a potential avenue for future investigation and improvement of the PLT method.

\subsection{Extended Lubrication Theory}\label{sec_ext_re}
An alternative extended lubrication theory model (T.G.-ELT) is proposed by Takeuchi \& Gu \cite{takeuchi_extended_2019}, which retains terms of order up to $\varepsilon^2$ in \cref{n-s_x_dimless_re,n-s_y_dimless_re} to result in the governing equations, 
 \begin{align}
\label{adj_gap_x} \frac{\partial P}{\partial X} &= \frac{\partial^2 U}{\partial Y^2} ,\\
\label{adj_gap_y} \frac{\partial P}{\partial Y} &= \varepsilon^2\frac{\partial^2 V}{\partial Y^2},
\end{align}
with incompressibility \cref{incompressibility}. Note that the second-order term $\varepsilon^2 \frac{\partial^2 U}{\partial X^2}$ appearing in \cref{n-s_x_dimless_re} is not included in the T.G.-ELT formulation, since they argue this term may be neglected in the case of small surface gradients \cite{takeuchi_extended_2019}. Moreover, the pressure gradient $\frac{\partial^2 P}{\partial Y\partial X}$ is considered negligible provided $\varepsilon^2 \ll 1$ \cite{takeuchi_extended_2019}. 

Since the $X$ momentum equations of lubrication theory and T.G.-ELT are the same and $\frac{\partial^2 P}{\partial Y\partial X}$ is considered negligible, integration of \cref{adj_gap_x} with the boundary conditions \cref{reyn_bc_u} gives $U=U_0$ as in \cref{reyn_u}. Then applying incompressibility and integrating with the boundary conditions \cref{reyn_bc_v} gives $V=V_0$ as in \cref{reyn_v}. However, the $Y$ momentum equations are different; integration of \cref{adj_gap_y} gives the pressure decomposition, 
\begin{equation} \label{tg_p} P(X,Y)=P_0(X)+ P_{\text{Adj}}(X,Y)+\varsigma(X),\end{equation} where $P_\text{Adj}(x,y) =\varepsilon^2 \frac{\partial V_0}{\partial Y}$ and $\varsigma(X)$ is an integral constant. Differentiating $V_0$ from \cref{reyn_v} yields,
\begin{equation} \label{adj_p} P_\text{Adj}(x,y) = \varepsilon^2\Bigg(\frac{-1}{2} \frac{d^2 P_0}{d X^2} Y^2 +\Bigg(\frac{1}{2}\bigg(\frac{d^2 P_0}{d X^2}H + \frac{dP_0}{d X}\frac{dH}{dX}\bigg) -\frac{U_b}{H^2}\frac{dH}{dX}\Bigg)Y\Bigg). \end{equation}

We assume the boundary condition \cref{reyn_bc_p} for $P_0$, and $P(1,0) = 0$. As with the PLT solution, note that the higher-order approximation $P$ does not necessarily satisfy the pressure gradient condition in \cref{reyn_bc_p}. Because the velocity is assumed to satisfy $U\equiv U_0$, the flux $\mathcal{Q}$ is the same as that of the order $\varepsilon^0$ solution. The average pressure drop is determined by the solution $P$, and in general $\Delta P_\text{T.G.-ELT}$ will differ from $\Delta P_0$. In \cite{takeuchi_extended_2019}, the integral constant $\varsigma$ is assumed negligible on account of $\varepsilon^2\ll 1$ and the assumption of small surface gradients. For $\varsigma = 0$, the average pressure drop is given by
$\Delta P_\text{ T.G.-ELT} 
    = \Delta P_0  +  \Delta P_\text{Adj}$. Since $P_\text{Adj}$ is polynomial in $Y$, $\Delta P_\text{Adj}$ may be evaluated exactly from \cref{dP} for a given $H$.

\section{Velocity-Adjusted Extended Lubrication Theory}
In the approach of Takeuchi \& Gu \cite{takeuchi_extended_2019}, the integral constant $\varsigma$ is assumed negligible. In the construction of the velocity to follow, we replace this assumption by requiring $\varsigma$ be chosen to ensure the velocity satisfies the boundary conditions and incompressibility, regardless of whether the T.G.-ELT conditions are satisfied. Including the potentially non-zero term $\varsigma$ leads to additional terms in $P$, $U$, and $V$ compared with the T.G.-ELT solution. We refer to the new solution including $\varsigma$ as the velocity adjusted extended lubrication theory solution (VA-ELT).

With $P$ of the form \cref{tg_p} and $P_\text{Adj}$ given by \cref{adj_p}, the $X$ momentum equation \cref{adj_gap_x} may be integrated exactly in $Y$ to obtain $U$. We utilize the integral constant $\varsigma$ to impose the boundary conditions and incompressibility independent of the local geometry. Applying the boundary conditions \cref{reyn_bc_u} for $U$ gives, \begin{multline} \label{adj_u} U(X,Y)= 
 \varepsilon^2\Bigg(\frac{- 1}{24}\frac{d^3P_0}{dX^3}(Y^4-H^3Y) + \frac{1}{12}\frac{d^2}{dX^2}\bigg[H\frac{dP_0}{dX}\bigg](Y^3-H^2Y)
\\+\frac{U_b}{6}\frac{d^2}{dX^2}\bigg[\frac{1}{H}\bigg](Y^3-H^2Y)\Bigg)+ \frac{1}{2}\frac{d\varsigma}{dX}(Y^2-HY)+U_0(X,Y).
\end{multline}
Then applying incompressibility \cref{incompressibility} to $U$ in \cref{adj_u}, and integrating with the boundary conditions \cref{reyn_bc_v} for $V$ gives, \begin{multline}\label{adj_v} V(X,Y)= \varepsilon^2\Bigg(\frac{1}{24}\bigg( \frac{d^4P_0}{dX^4}\bigg(\frac{Y^5}{5}-\frac{H^3Y^2}{2}\bigg)-\frac{d^3 P_0}{dX^3}\frac{dH}{dX}\frac{3H^2Y^2}{2} \bigg)\\
- \frac{1}{12}\bigg(\frac{d^3}{dX^3}\bigg[H\frac{dP_0}{dX}\bigg]\bigg(\frac{Y^4}{4}-\frac{H^2Y^2}{2}\bigg) -\frac{d^2}{dX^2}\bigg[H\frac{dP_0}{dX}\bigg]\frac{dH}{dX}HY^2\bigg)\\
+\frac{U_b}{6}\bigg(\frac{d^3H}{dX^3}\frac{Y^4-2H^2Y^2}{4H^2} + \bigg[\frac{dH}{dX}\bigg]^3\frac{3Y^4-2H^2Y^2}{2H^4}+\frac{dH}{dX}\frac{d^2H}{dX^2}\frac{4H^2Y^2-3Y^4}{2H^3}\bigg)\Bigg)\\
-\frac{d^2\varsigma}{dX^2}\bigg(\frac{Y^3}{6}-\frac{HY^2}{4}\bigg)+\frac{d\varsigma}{dX}\frac{dH}{dX}\frac{Y^2}{4} +V_0(X,Y).\end{multline}
Observe that if we assume $\varsigma=0$, the velocity field from \cref{adj_u,adj_v} is not necessarily incompressible, and \cref{adj_v} may not satisfy $V(X,H)=0$.

The pressure $P_0$ is assumed to satisfy the boundary conditions \cref{reyn_bc_p}. For the higher-order pressure, we assume $P(1,0)=0$, but note that $P$ does not necessarily satisfy the pressure gradient condition in \cref{reyn_bc_p}. The integral constant $\varsigma$ is determined by requiring a constant flux, and as a consequence the average pressure drop is determined by \cref{dP}. We prescribe $\mathcal{Q}$ to be consistent with the flux of the order $\varepsilon^0$ solution. Then $\varsigma$ satisfies,
\begin{multline}\frac{d\varsigma}{dX}= \varepsilon^2\Bigg(\frac{3H^2}{20}\frac{d^3P_0}{dX^3} -\frac{H}{4}\bigg(H\frac{d^3P_0}{dX^3}+2\frac{d^2P_0}{dX^2}\frac{dH}{dX}+\frac{dP_0}{dX}\frac{d^2H}{dX^2}\bigg)
\\+\frac{U_b}{2}\bigg(\frac{1}{H}\frac{d^2H}{dX^2}-\frac{2}{H^2}\bigg[\frac{dH}{dX}\bigg]^2 \bigg)\Bigg).\end{multline}
For an arbitrary height function, numerical integration is required to obtain $\varsigma$. The boundary condition $P(1,0)=0$ corresponds to $\varsigma(1)=0$. The average pressure drop is then
$\Delta P_\text{ VA-ELT} 
    = \Delta P_0  + \Delta P_\text{Adj} + \varsigma(0)$.

\section{Results}\label{sec_comparing}
We now compare the lubrication models for various geometries and examine the solutions of velocity and pressure for a variety of textured slider examples. The Reynolds, T.G.-ELT, VA-ELT, $\varepsilon^2$-PLT and $\varepsilon^4$-PLT solutions are compared against the Stokes solution (\cref{app_stokes}). The accuracy of the pressure and velocity solutions is characterized using the relative percent error in $L_2$ norm. For an appropriate comparison of solutions from the various models, the flux $\mathcal{Q}$ is kept constant and the average pressure drops $\Delta P$ are presumed to vary. 

Note, for the T.G.-ELT method, the prescribed flux $\mathcal{Q}$ may not be conserved throughout the domain – subject to the magnitude of surface variation – because incompressibility is not strictly enforced.

\subsection{The logistic step}\label{sec_logistic}
\begin{figure}[h] 
 \centering 
 \includegraphics[width=.95\textwidth]{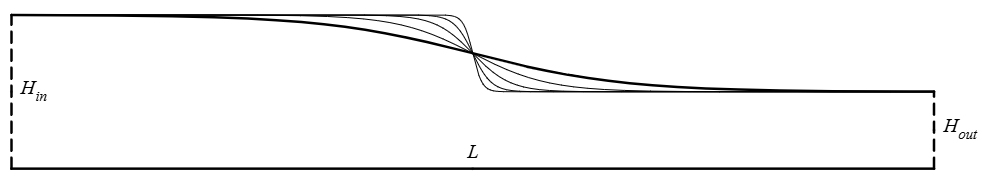}
 \caption{Schematic of the logistic step.}\label{schematic_logistic} 
\end{figure}

First, we consider the logistic step, shown in \cref{schematic_logistic}. The height function is given {\color{black} in dimensional terms} by, 
\begin{equation}\label{logistic_eq} h(x) = H_{\text{in}} + \frac{H_{\text{out}}-H_{\text{in}}}{1+e^{\lambda \big(L/2-x\big)}},\end{equation} 
where $H_\text{in} \ge H_\text{out}>0$ and $\lambda > 0$.
The logistic step allows us to vary the magnitude of surface gradient while keeping a fixed length scale ratio $\varepsilon = L_y/L_x$ and step expansion ratio $H_\text{in}/H_\text{out}$. We consider solutions for $2\le \lambda \le 32$, and with fixed $H_\text{in}=2$, $H_\text{out}=1$, $L=16$, $\nu=1$, $\mathcal{U}=0$ and $\mathcal{Q}=1$. Then the length scale ratio is $\varepsilon \simeq 1/8$, and the maximum magnitude of surface gradient is $\big|\frac{dh}{dx}\big| = \lambda(H_\text{in}-H_\text{out})/4=\lambda/4$.

\begin{figure}
    \centering
    \subfloat[Pressure contours $p(x,y)$]{
        \makebox[.95\textwidth]{
            \includegraphics[width=1.25\textwidth]{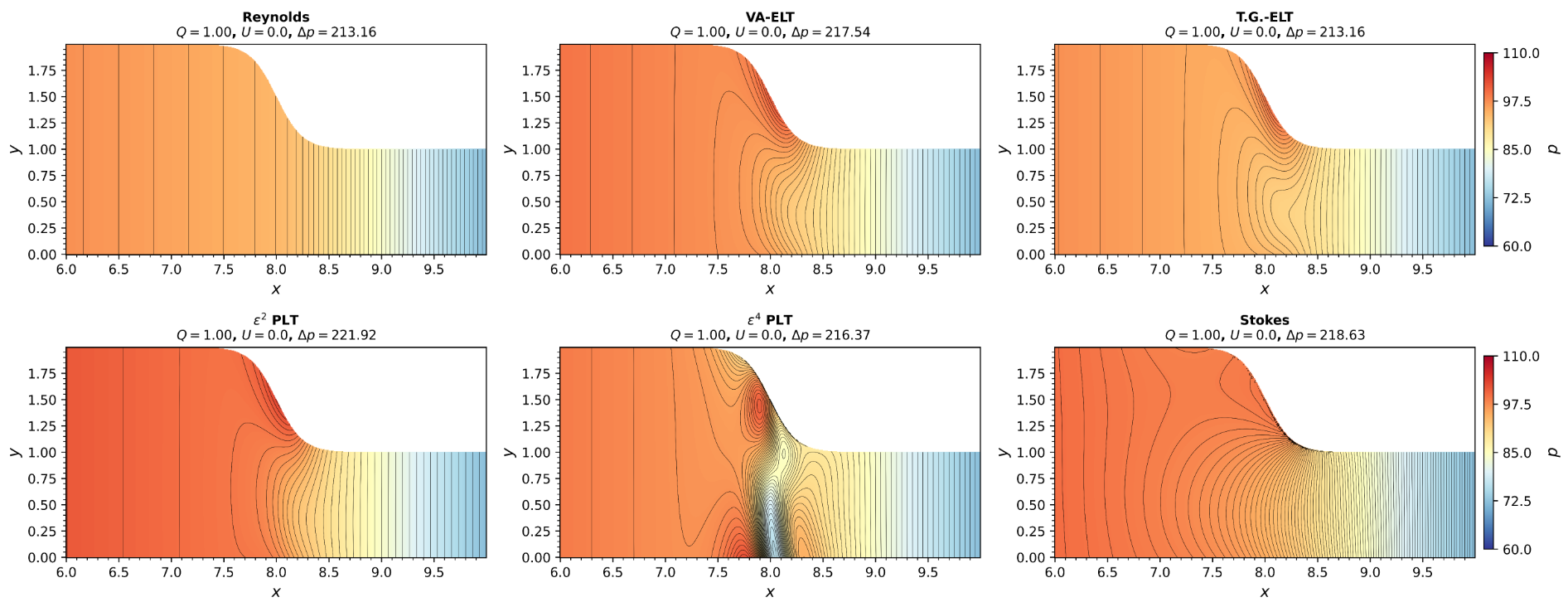}
        }
    }
    
    \subfloat[Velocity streamlines $(u,v)$]{
        \makebox[.95\textwidth]{
            \includegraphics[width=1.25\textwidth]{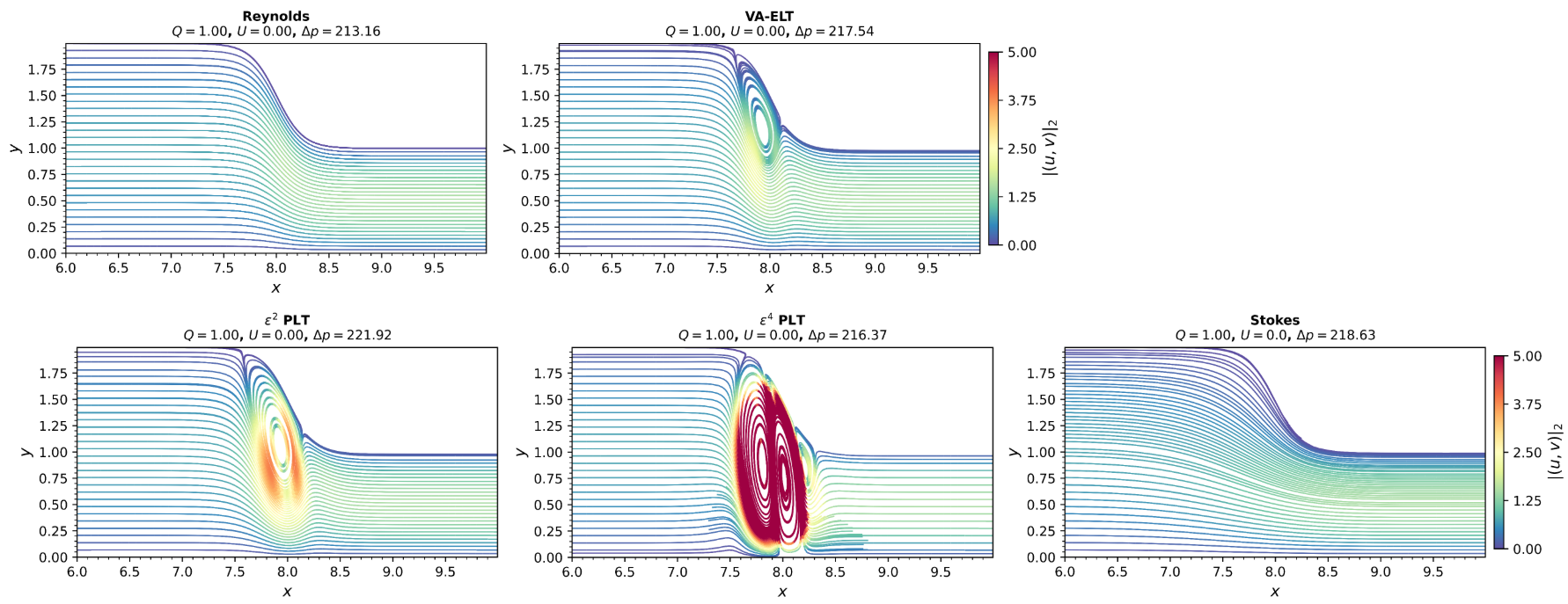}
        }
   }
   \caption{Pressure and velocity solutions for the logistic step with $\lambda = 8$. The ELT and $\varepsilon^k$-PLT solutions overestimate $\frac{\partial p}{\partial y}$ in the vicinity of the surface variation, leading to flow recirculation which is not observed in the Stokes solution at this moderate $\lambda$. The velocity for T.G.-ELT is not shown as it is equivalent to the Reynolds velocity.}\label{logistic_d8}
\end{figure}

The pressure and velocity for the logistic step with $\lambda=8$ are shown in \cref{logistic_d8}. The pressure contours for the Stokes solution exhibit significant $\frac{\partial p}{\partial y}$ in a wider region surrounding the large surface gradient compared with the ELT and $\varepsilon^k$-PLT solutions, whereas the ELT and $\varepsilon^k$-PLT solutions overestimate $\frac{\partial p}{\partial y}$ in the localized region of surface variation. This leads to flow recirculation being observed at smaller $\lambda$ in the ELT and $\varepsilon^k$-PLT solutions than with the Stokes solution.

\begin{figure}
    \centering
     \makebox[.95\textwidth]{
         \includegraphics[width=1.25\textwidth]{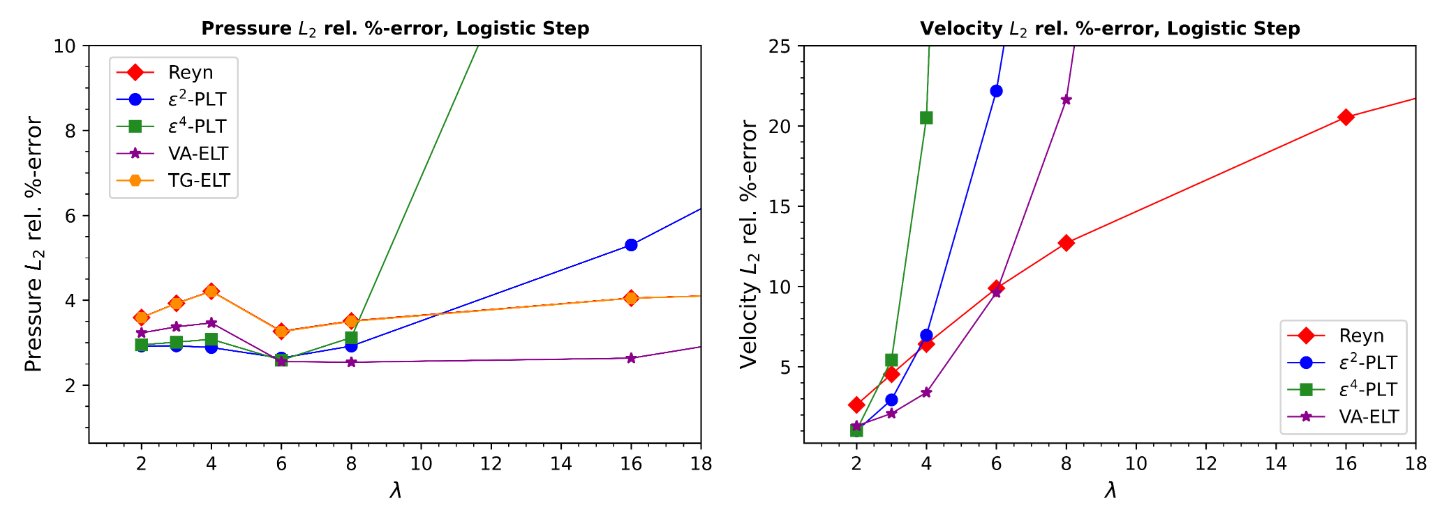}
    }
   \caption{Relative percent error in velocity $(u,v)$ and pressure $p(x,y)$ compared with the Stokes solution for the logistic step at varying slopes, $\lambda/4 = \max |\frac{dh}{dx}|$. For $2 < \lambda \le 6$, the VA-ELT solution has the smallest error in velocity and the $\varepsilon^2$-PLT solution has the smallest error in pressure.}\label{logistic_errs}
\end{figure}

The relative percent errors in $l_2$ norm of velocity and pressure compared with the Stokes solution are shown in \cref{logistic_errs}. For the wide range of $\lambda < 32$, the VA-ELT solution of pressure is significantly more accurate than the Reynolds and T.G.-ELT solutions, and the T.G.-ELT solution of pressure yields a marginal improvement over the Reynolds pressure.  For $2 < \lambda < 8$, the $\varepsilon^2$-PLT solution is best at approximating the pressure, and the VA-ELT solution is best at approximating the velocity. The $\varepsilon^2$-PLT solution performs better than the $\varepsilon^4$-PLT solution for all $\lambda > 2$. For $\lambda \ge 32$, the Reynolds solution has the smallest error in velocity and pressure; the Reynolds solution has the most stable error across the range of $\lambda$. Overall, both the $\varepsilon^2$-PLT and VA-ELT solutions perform well for the logistic step up to moderate surface variation. We note however, that all the lubrication models discussed here break down in the limit $\lambda \to \infty$ of a discontinuous step, see \cref{app_BFS}.

\subsection{The triangular slider}\label{sec_tri}
\begin{figure}[h] 
 \centering 
 \includegraphics[width=.95\textwidth]{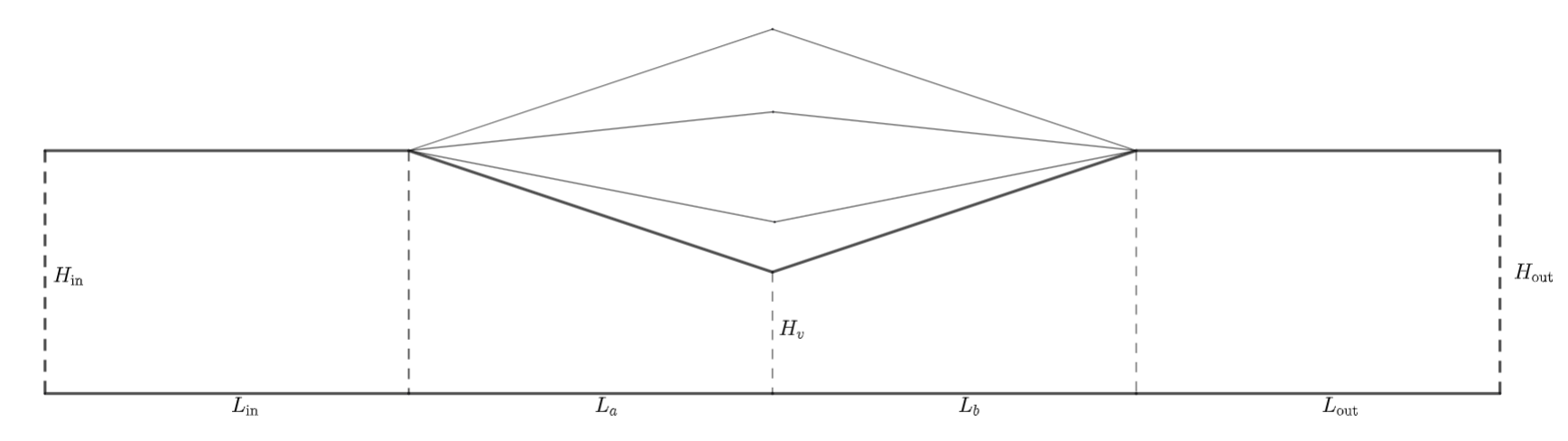}
 \caption{Schematic of the triangular slider} \label{schematic_tri}
\end{figure}

Next, we consider a triangular slider, shown in \cref{schematic_tri}.  The height function is given {\color{black} in dimensional terms by},
\begin{equation}\label{trislider_eq}
h(x)=\begin{cases}
H_\text{in} & x \le L_\text{in}\\
H_v - \frac{H_v-H_\text{in}}{L_a}\big(L_\text{in}+ L_a -x\big) & L_\text{in}\le x\le  L_\text{in} + L_a\\
    H_\text{out} - \frac{H_\text{out}-H_v}{L_b}\big(L-L_\text{out} - x\big) &L_\text{in} + L_a \le x \le L-L_\text{out}\\
    H_\text{out} &  L-L_\text{out}\le x \le  L.
\end{cases}\end{equation}
We consider varying central vertex heights $\frac{1}{16}\le H_v\le 2$, and with fixed $L_a=1.25$, $L_b=0.75$, $L_\text{in} = L_\text{out} = 7$, $H_\text{in}=H_\text{out} = 1$, $\nu=1$, $\mathcal{U}=0$ and $\mathcal{Q}=1$. The maximum magnitude of surface gradient is $|\frac{dh}{dx}|=|1-H_v|/0.75$, and the length scale ratio is $\varepsilon \simeq \max(1,H_v)/16$. This configuration is similarly considered in \cite{tavakol_extended_2017} for the case of negative texturing $(H_v< 1)$ with the $\varepsilon^k$-PLT solutions. 

\begin{figure}
    \centering
    \subfloat[Pressure contours $p(x,y)$]{
    \makebox[.95\textwidth]{
            \includegraphics[width=1.25\textwidth]{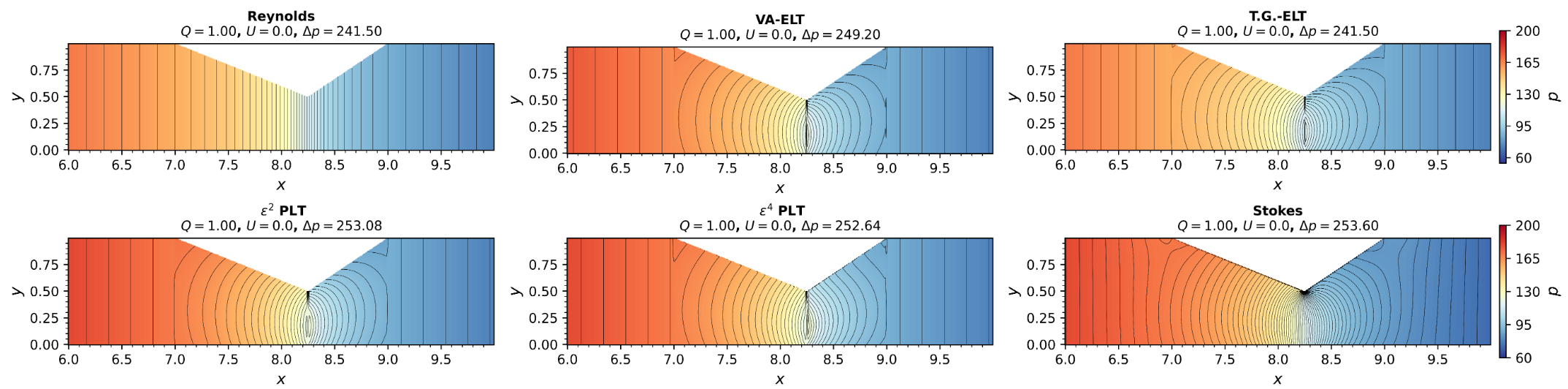}
        }}
    
    \subfloat[Velocity streamlines $(u,v)$]{
        \makebox[.95\textwidth]{
            \includegraphics[width=1.25\textwidth]{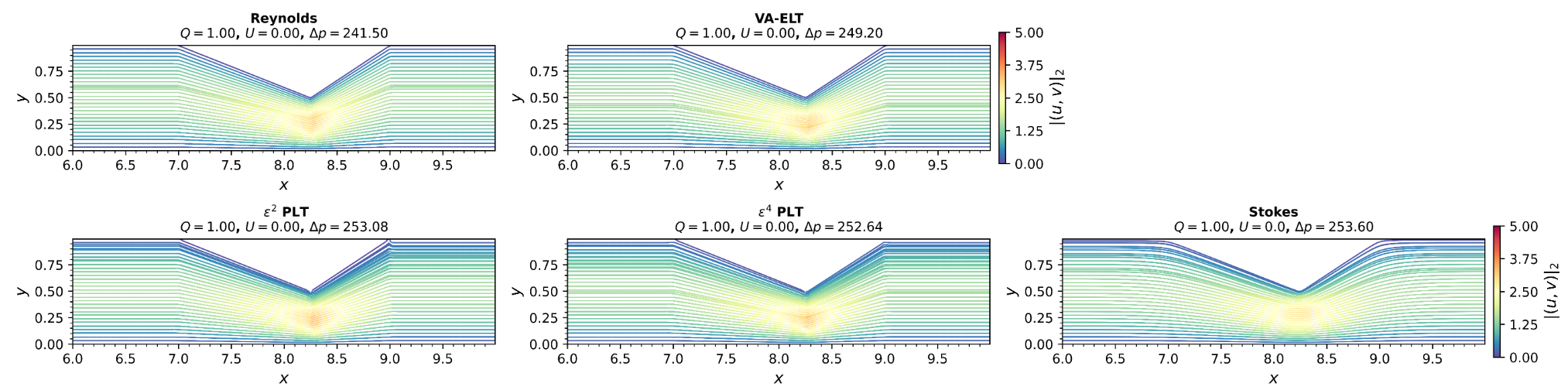}
    }}
    \caption{Pressure and velocity solutions for the triangular slider with $H_v=0.5$. The Reynolds, ELT and PLT models  overestimate the maximum velocity. Discontinuities in velocity and pressure of the ELT and $\varepsilon^k$-PLT models are less pronounced at smaller $H_v$.}\label{trislider_h0.5}
\end{figure}

\begin{figure}
    \centering
    \subfloat[Pressure contours $p(x,y)$]{
        \makebox[.95\textwidth]{
            \includegraphics[width=1.25\textwidth]{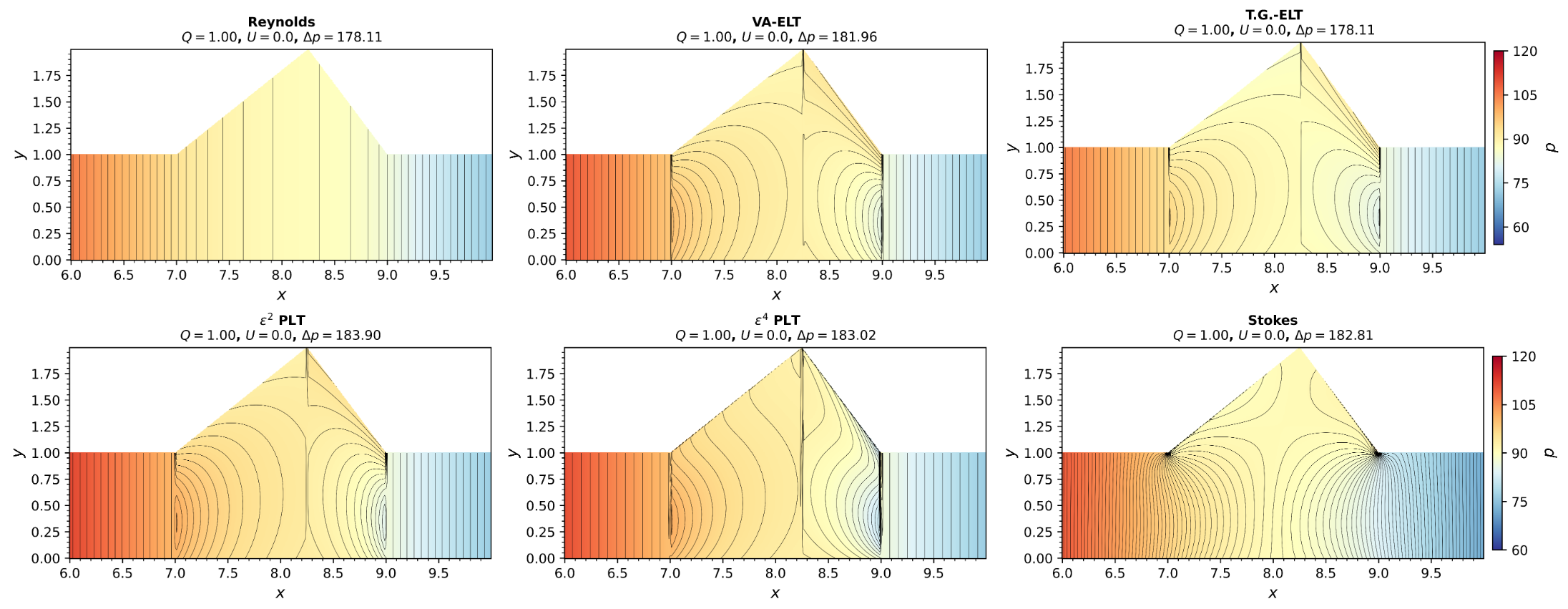}
    }}
    
    \subfloat[Velocity streamlines $(u,v)$]{
        \makebox[.95\textwidth]{
            \includegraphics[width=1.25\textwidth]{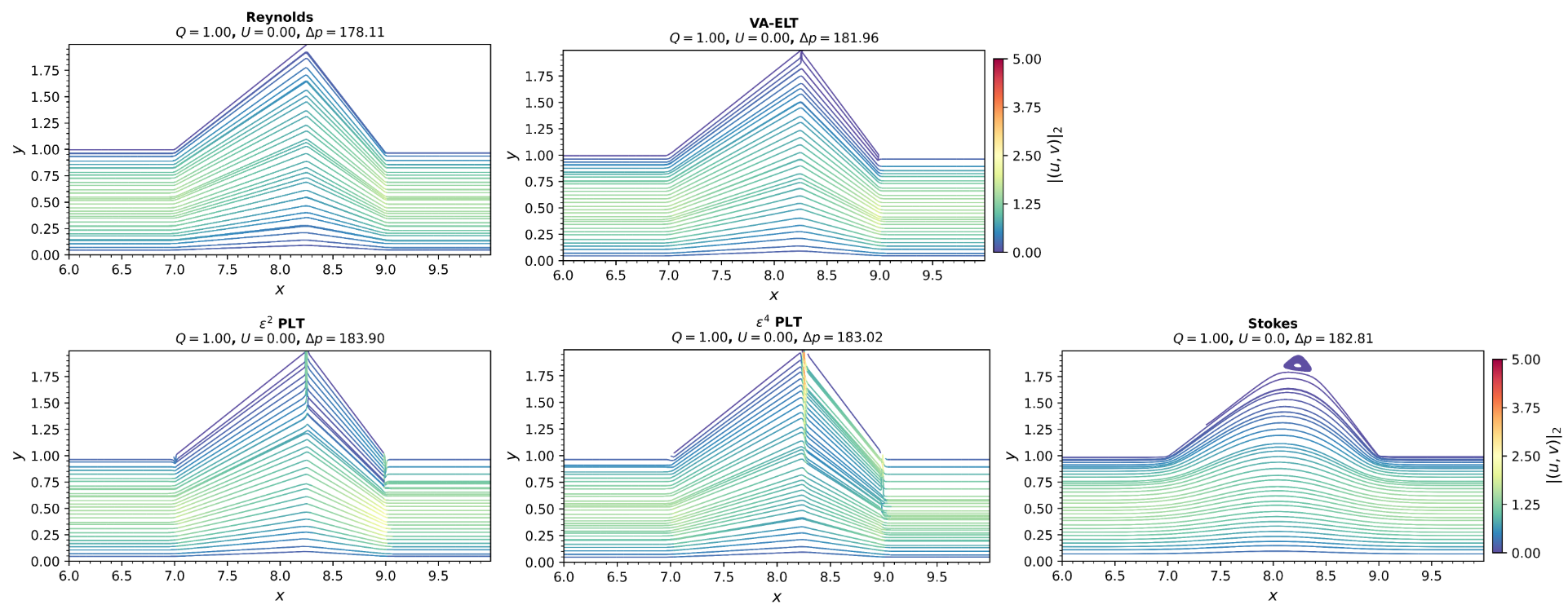}
    }}
    \caption{Pressure and velocity solutions for the triangular slider with $H_v=2$. The velocity and pressure solutions of ELT and $\varepsilon^k$-PLT are discontinuous at the discontinuities of $\frac{dh}{dx}$. The sequence of corner eddies characteristic of the Stokes solution is not observed in any of the lubrication models.}\label{trislider_h2}
\end{figure}

The pressure and velocity solutions to the triangular slider with $H_v=0.5$ (following \cite{tavakol_extended_2017}) and $H_v=2$ are shown in \cref{trislider_h0.5,trislider_h2} respectively. This class of examples demonstrates the limitations of the lubrication models in the case of a discontinuous surface gradient. At the discontinuities in $\frac{dh}{dx}$, the pressure gradient $\frac{d^2p_0}{dx^2}$ and velocity $v_0$ of the Reynolds solution are not necessarily continuous. Consequently, the pressure and velocity solutions from ELT and $\varepsilon^k$-PLT may not be continuous at the discontinuities of $\frac{dh}{dx}$. Note that $u_0$ and $\frac{dp_0}{dx}$ are continuous for the triangular slider because $h(x)$ is continuous; owing to the lubrication assumptions, $v_0$ is typically small enough in magnitude compared to $u_0$ that the discontinuity in $v_0$ is not significant. However, for larger $H_v$, the discontinuity in $v_0$ is more pronounced and leads to a noticeable loss of smoothness in the ELT and $\varepsilon^k$-PLT solutions. 

In the case of a sufficiently small interior angle at the triangle apex, the Stokes solution is known to exhibit a sequence of eddies receding into the corner \cite{moffatt_viscous_1964,biswas_hoc_2017}. For $H_v=2$ as in \cref{trislider_h2}, the first eddy is clearly observed in the Stokes solution. For the ELT and $\varepsilon^k$-PLT solutions, flow recirculation may occur if $H_v$ and $\frac{dh}{dx}$ are sufficiently large. However, because the velocities of these models are not continuous at the discontinuities of $\frac{dh}{dx}$, corner eddies do not occur as in the Stokes solution. In fact, as seen in \cref{trislider_h2} with the $\varepsilon^k$-PLT solutions, the velocity in the steeper half of the triangle texture exhibits flow recirculation at smaller $H_v$ than the shallower half. That is, if $H_v$ is large enough and $\frac{dh}{dx}$ is discontinuous in magnitude, the Reynolds, ELT and $\varepsilon^k$-PLT solutions may exhibit spontaneous flow reversal at the discontinuities of $\frac{dh}{dx}$.

\begin{figure}
    \centering
     \makebox[.95\textwidth]{
     \includegraphics[width=1.25\textwidth]{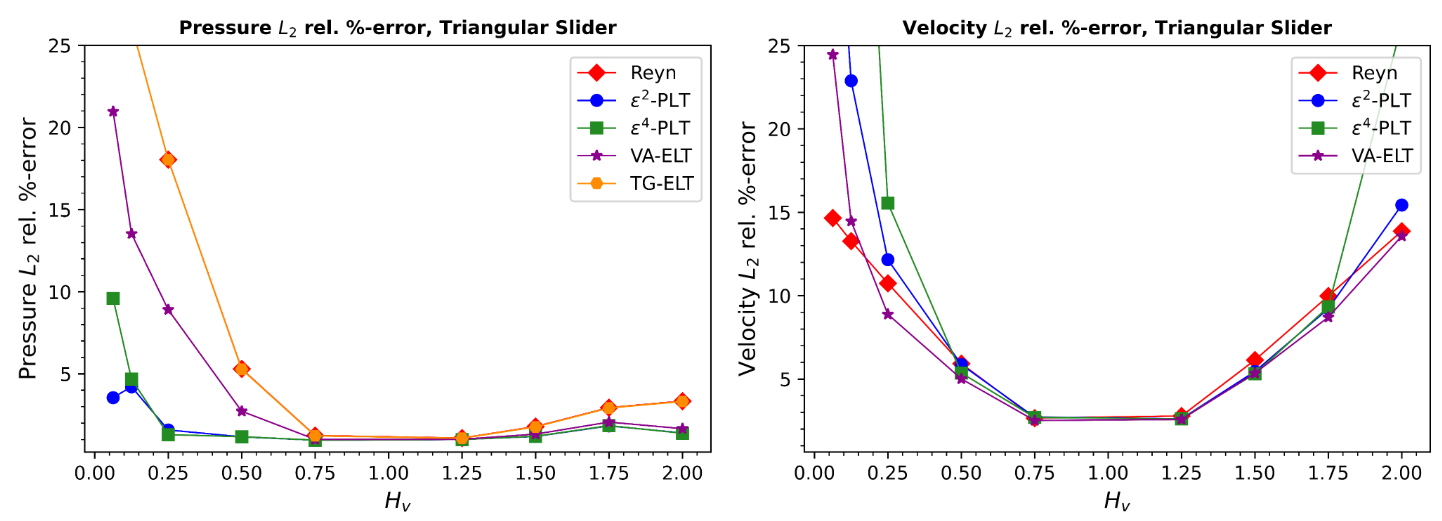}
    }
    
   \caption{Relative percent error in velocity $(u,v)$ and pressure $p(x,y)$ compared with the Stokes solution for the triangular slider with varying apex height $H_v$. Decreased difference between maximum and minimum height corresponds to decreased error in pressure and velocity.}\label{trislider_errs}
\end{figure}

The relative percent errors in the $l_2$ norm of pressure and velocity compared with the Stokes solution are shown in \cref{trislider_errs}. The error in velocity is very similar for all of the models. For pressure, the VA-ELT solution is a consistent improvement compared to the Reynolds and T.G.-ELT solutions, and the $\varepsilon^k$-PLT solutions perform better still. For negative texturing ($H_v < 1$), the $\varepsilon^k$-PLT solutions of pressure have significantly smaller error relative to both the VA-ELT and Reynolds solutions; for positive texturing ($H_v > 1$), the $\varepsilon^k$-PLT and VA-ELT solutions are more similar. The error in both pressure and velocity decreases as the difference between the minimum and maximum height decreases. In particular, for the case of negative texturing, the increasing magnitude of surface gradient as the minimum height decreases corresponds to an increase in error for pressure and velocity; this emphasizes the importance of considering the magnitude of surface gradients as well as the length scale ratio when selecting a fluid model.

\section{Conclusions}
We considered several models in extended lubrication theory and compared the solutions of velocity and pressure across a range of geometries, including several configurations in which the thin film assumption of classical lubrication theory is not strictly satisfied. The VA-ELT solution proposed here amends the velocity and pressure of the T.G.-ELT solution to ensure that incompressibility and the boundary conditions are satisfied. To assess the sensitivity of the lubrication models to the thin film assumption and to large magnitudes of surface variation in the low Reynolds number limit, the Reynolds, VA-ELT, T.G.-ELT, $\varepsilon^2$-PLT and $\varepsilon^4$-PLT solutions were compared with the Stokes solution for the logistic step, the backward facing step, and the triangular slider with positive and negative texturing.

We found that all lubrication models have increased error in pressure and velocity as the magnitude of surface variation increases, while the length scale ratio $\varepsilon = L_y/L_x$ is kept constant. In the presence of a large surface gradient, the Stokes solution exhibits significant $\frac{\partial p}{\partial y}$ in a region surrounding the surface gradient, compared with the ELT and $\varepsilon^k$-PLT solutions which localize this pressure variation to regions where the surface gradient is large. The consequences of this localization is overestimation of $\frac{\partial p}{\partial y}$, overestimation of the velocity magnitude, and in the case of very large surface gradients, spurious flow recirculation.

For the logistic step, the VA-ELT and $\varepsilon^k$-PLT solutions are an improvement on the Reynolds solution for smaller slopes; but at larger slopes, particularly for $\varepsilon^k$-PLT, the error exceeds that of the Reynolds solution. For smaller slopes, the VA-ELT solution has the smallest error in velocity, and the $\varepsilon^2$-PLT solution has the smallest error in pressure. At larger slopes, the VA-ELT and $\varepsilon^k$-PLT solutions exhibit flow recirculation at too small of slope compared with the Stokes solution, and the velocity magnitude in the recirculation region becomes unreasonably large given the small Reynolds number.  

For the triangular slider, the error in velocity is very similar for all models. For pressure, the $\varepsilon^k$-PLT solutions have significant improvements in error, particularly for the case of negative texturing. In the presence of the same magnitude of surface variation but at positive texturing, the VA-ELT model performs similarly to the $\varepsilon^k$-PLT models. This suggests that for small heights, the $\varepsilon^k$-PLT models are less sensitive to large or discontinuous surface gradients than the VA-ELT model. However, both the ELT and $\varepsilon^k$-PLT models exhibit discontinuities when the surface gradient is discontinuous. Furthermore, the sequence of corner eddies characteristic of the Stokes solution is not observed in the lubrication models. 

These textured slider examples demonstrate the importance of considering both the magnitude of surface gradient and the length scale ratio when selecting a reduced order model for a low Reynolds number fluid. In the presence of only small surface gradients, significant improvements on the Reynolds solution can be achieved with an extended or perturbed model of lubrication theory; however, as demonstrated in the examples explored here, these models can be sensitive to the surface gradient.

\section*{Acknowledgments}We thank Michael Fillon for providing feedback on an early version of this manuscript. This work was supported by National Science Foundation (NSF) grant DMS-2512565 to TGF. We also acknowledge use of the Brandeis High Performance Computing Cluster (HPCC) which is partially supported by the NSF through DMR-MRSEC 2011846 and OAC-1920147.

\appendix

\section{The Reynolds equation finite difference solution}\label{app_reyn_fd}
Define the uniform discretisation of $[0,1]$,
\begin{align}\{X_i\}_{i=0}^N, && X_i = i \Delta X, && \Delta X = \tfrac{1}{N}.
\end{align}
A second-order accurate difference approximation for the Reynolds equation \cref{reynolds} is,
\begin{multline}\label{reyn_disc}
\Big(H_{i+1}^3 + H_i^3\Big)P_{i+1} - \Big(H_{i+1}^3 + 2H_i^3 +H_{i-1}^3\Big)P_i + \Big(H_i^3 +H_{i-1}^3\Big)P_{i-1}\\= 6 U_b \Big(H_{i+1}-H_{i-1}\Big)\Delta X.
 \end{multline}

\section{The Stokes Equation} \label{app_stokes}
The Stokes equation is derived from the Navier-Stokes equation under the assumption of zero Reynolds number and a steady state flow \cite{leal_advanced_2007}. Compared with the lubrication approximation, Stokes flow is formulated with a single characteristic velocity $U_*$ and length scale $L_x$. The characteristic pressure is now $\tilde P_* = \eta U_*/L_x$. Consequently, Stokes flow does not involve the length scale ratio $\varepsilon$ and is not sensitive to surface variation as the lubrication models are.

With the dimensionless variables $X = x/L_x$, $Y = y/L_x$, $U = u/U_*$, $V = v/U_*$, $T = t/T_*$ and $P = p/\tilde P_*$, the Navier-Stokes equations \cref{n-s_x,n-s_y} may be expressed as 
\begin{align}\label{n-s_x_dimless_Stokes}
 \frac{\partial P}{\partial X} &= \Big(\frac{\partial^2 U}{\partial X^2} + \frac{\partial^2 U}{\partial Y^2}\Big) - \text{Re}\Big( \frac{\partial U}{\partial T}+U\frac{\partial U}{\partial X} + V \frac{\partial U}{\partial Y} \Big),\\
 \label{n-s_y_dimless_Stokes} \frac{\partial P}{\partial Y} &=  \Big(\frac{\partial^2 V}{\partial X^2} + \frac{\partial^2 V}{\partial Y^2}\Big) - \text{Re}\Big(  \frac{\partial V}{\partial T}+U\frac{\partial V}{\partial X} + V \frac{\partial V}{\partial Y}\Big).
\end{align}
Through introduction of the stream function $\Psi(X,Y)$ satisfying,
\begin{align} \label{stream}
 \hfill&& U = \frac{\partial\Psi}{\partial Y} ,\hspace{3em} V = -\frac{\partial \Psi}{\partial X},&&\hfill
\end{align}
and assuming a steady state flow, the equations \cref{n-s_x_dimless_Stokes,n-s_y_dimless_Stokes}
may be expressed as,
\begin{equation}\label{n-s_psivel}
 \nabla^4\Psi = \text{Re}\big(V\nabla^2 U - U \nabla^2 V).\end{equation}
The formulation \cref{n-s_psivel,stream} presents an effective method of solution to the Navier-Stokes equations for low to moderate Reynolds numbers \cite{gupta_new_2005,marner_potential-velocity_2014,sen_4oec_2015,biswas_hoc_2017}. Here, we restrict to $\text{Re}=0$, in which case \cref{n-s_psivel} reduces to the Stokes equation $\nabla^4 \Psi=0$. 

To solve the Stokes equation, we assume the inlet and outlet flow profiles correspond to a fully developed laminar flow with flux $\mathcal{Q}$. The surface boundary conditions for velocity are \cref{reyn_bc_u,reyn_bc_v}, and the pressure satisfies \cref{reyn_bc_p}. The inlet and outlet velocity profiles are,
\begin{align}\label{n-s_bc_vel}
 U(0,Y) &= U_0(0,Y) &&\frac{\partial U}{\partial X}\bigg|_{1,Y} = 0,\\
 V(0,Y) &= 0 && V(1,Y) = 0.
\end{align}
The corresponding boundary conditions for the stream function $\Psi$ are,
\begin{align}\label{n-s_bc_stream}
 \Psi(0,Y) &= \int_{0}^Y U_0(X,\hat{Y})d\hat {Y} &&\frac{\partial \Psi}{\partial X}\Big|_{1,Y} = 0,\\
 \Psi(X, 0) &= 0 & &\Psi(X,H) = 1,
\end{align}
where,
\begin{equation} \label{n-s_bc_stream_x0}
\int_{0}^Y U_0(X,\hat{Y})d\hat{Y} =\frac{Y^2}{H^3}\Big(3H-2Y\Big) +  \frac{U_bY}{H^2}\Big(H-Y\Big)^2.
\end{equation}

The solution $\Psi$, $U$, $V$ to the Stokes equation is determined through an iterative second-order accurate finite difference method. The method presented here is adapted from the method of \cite{biswas_hoc_2017,gupta_new_2005} for the full biharmonic Navier-Stokes equations \cref{n-s_psivel,stream}. The Stokes equation $\nabla^4 \Psi=0$ is discretized as,
\begin{multline}\label{n-s_disc}
    28\Psi_{i,j} -  8  \big(\Psi_{i-1,j}+\Psi_{i+1,j}+\Psi_{i,j-1}+\Psi_{i,j+1}\big) \\ + \big(\Psi_{i-1,j-1}+\Psi_{i-1,j+1}+\Psi_{i+1,j-1}+\Psi_{i+1,j+1}\big)
     \\ = 3\Delta X \big(U_{i,j-1} - U_{i,j+1} + V_{i+1,j}-V_{i-1,j}\big)
\end{multline}
The stream-velocity relation \eqref{stream} is discretised as,
\begin{align}\label{n-s-vel_disc_u}
    U_{i,j} &= \frac{-3}{4\Delta X}\big(\Psi_{i,j-1}-\Psi_{i,j+1}\big) - \frac{1}{4}\big(U_{i,j-1}+U_{i,j+1}\big) 
    \\ \label{n-s-vel_disc_v}
    V_{i,j} &= \frac{3}{4\Delta X}\big(\Psi_{i-1,j}-\Psi_{i+1,j}\big)-\frac{1}{4}\big(V_{i-1,j}+V_{i+1,j}\big).
\end{align}
In \cite{dennis_separation_2026}, we present a method for approximating the stream and velocity boundary conditions on non-rectilinear domains. 

Once the stream and velocity have sufficiently converged, the pressure partial derivatives are determined using a centered second-order accurate finite difference discretisation of the dimensionless Navier-Stokes equations \cref{n-s_x_dimless_Stokes,n-s_y_dimless_Stokes} at $\text{Re}=0$. For \cref{n-s_x_dimless_Stokes},
\begin{equation}\label{n-s-x_disc}\frac{\partial P}{\partial X}\Big|_{i,j} = \frac{1}{\Delta X^2} \Big(U_{i-1,j} +U_{i+1,j} - 4U_{i,j} + U_{i,j-1} + U_{i,j+1}\Big), 
\end{equation} and $\frac{\partial P}{\partial Y}$ as in \cref{n-s_y_dimless_Stokes} is similar. The pressure $P(X,Y)$ is then determined numerically by a path integral from the outlet with the boundary condition $P(1,0)=0$. The pressure drop $\Delta P_\text{Stokes}$ is determined by numerical integration according to \cref{dP}.

\section{Perturbed Lubrication Theory}\label{app_pert}
The dimensionless expressions for $P_k$, $U_k$, and $V_k$ at orders $\varepsilon^2$ and $\varepsilon^4$ are provided. 
For order $\varepsilon^2$,\begin{equation} P_2(X,Y)=-\frac{\partial U_0}{\partial X} + \int_0^{X} \frac{d\gamma_3}{d\hat{X}}d\hat{X},\end{equation}
\begin{equation}U_2(X,Y) =  \frac{d^2\big[H^{-3}\big]}{dX^2}\Big(Y^4 - H^3Y\Big) -2\frac{d^2\big[H^{-2}\big]}{dX^2}\Big(Y^3 - H^2Y\Big) + \frac{1}{2}\frac{d\gamma_3}{dX}(Y^2 -HY),\end{equation}
and,
\begin{multline} \label{pert_O2_v}V_2(X,Y) = - \frac{d^3\big[H^{-3}\big]}{dX^3}\Big(\tfrac{1}{5}Y^5 - \tfrac{1}{2}H^3 Y^2\Big) + \frac{3}{2}\frac{d^2\big[H^{-3}\big]}{dX^2}\frac{dH}{dX}H^2Y^2 \\+\frac{d^3\big[H^{-2}\big]}{dX^3}\Big(\tfrac{1}{2}Y^4 - H^2Y^2\Big) - 2\frac{d^2\big[H^{-2}\big]}{dX^2}\frac{dH}{dX}HY^2 \\- \frac{1}{2}\frac{d^2 \gamma_3}{dX^2}\Big(\tfrac{1}{3}Y^3 - \tfrac{1}{2}HY^2\Big) + \frac{1}{4}\frac{d\gamma_3}{dX}\frac{dH}{dX}Y^2,
\end{multline}
where,
\begin{equation}
\frac{d \gamma_3}{dX} =  - \frac{18}{5}\frac{d^2\big[H^{-3}\big]}{dX^2}H^2+6\frac{d^2\big[H^{-2}\big]}{dX^2}H.
\end{equation}
And likewise for order $\varepsilon^4$,
\begin{equation}\label{pert_O4_p}P_4(X,Y) = -\frac{\partial U_2}{\partial X} +\int_0^{Y}\frac{\partial^2 V_0 }{\partial {X^2}}  d\hat{Y} + \int_0^{X} \frac{d\gamma_5}{d\hat{X}}d\hat{X},\end{equation}
\begin{multline}
    U_4(X,Y)=  -\frac{1}{20}\frac{d^4\big[H^{-3}\big]}{dX^4}\Big(Y^6-H^5Y\Big) + \frac{3}{20}\frac{d^4\big[H^{-2}]}{dX^4}\Big(Y^5-H^4Y\Big) \\
   + \Bigg(\frac{1}{3}\frac{d^2\phi_1}{dX^2}-\frac{2}{3}\frac{d^2\phi_2}{dX^2} + \frac{1}{6}\frac{d^2\phi_3}{dX^2}\Bigg)\Big(Y^3-H^2Y\Big) \\ 
    -\frac{1}{12}\frac{d^3 \gamma_3}{dX^3}\Big(Y^4-H^3Y\Big) + \frac{1}{2}\frac{d\gamma_5}{dX}\Big(Y^2-HY\Big),
\end{multline}
and,
\begin{multline}
    V_4(X,Y) = -\frac{1}{20}\frac{d^5\big[H^{-3}\big]}{dX^5}\Big(\tfrac{1}{7}Y^7 - \tfrac{1}{2}H^5Y^2\Big) + \frac{1}{8}\frac{d^4\big[H^{-3}\big]}{dX^4}\frac{dH}{dX}H^4Y^2 \\
    + \frac{3}{20}\frac{d^5\big[H^{-2}\big]}{dX^5}\Big(\tfrac{1}{6}Y^6-\tfrac{1}{2}H^4Y^2\Big) - \frac{3}{10}\frac{d^4\big[H^{-2}\big]}{dX^4}\frac{dH}{dX}H^3Y^2 \\
    +\frac{1}{3}\Bigg(\frac{d^3\phi_1}{dX^3} -2\frac{d^3\phi_2}{dX^3} \Bigg)\Big(\tfrac{1}{4}Y^4-\tfrac{1}{2}H^2Y^2\Big)
    -\Bigg(\frac{d^2\phi_1}{dX^2} -2\frac{d^2\phi_2}{dX^2} \Bigg)\frac{dH}{dX}HY^2\\
    +\frac{1}{6}\Bigg(\frac{d^3\phi_3}{dX^3}\Big(\tfrac{1}{4}Y^4-\tfrac{1}{2}H^2Y^2\Big)-\frac{d^2\phi_3}{dX^2}\frac{dH}{dX}HY^2\Bigg) -\frac{1}{12}\frac{d^4\gamma_3}{dX^4}\Big(\tfrac{1}{5}Y^5-\tfrac{1}{2}H^3Y^2\Big) \\
    + \frac{1}{8}\frac{d^3\gamma_3}{dX^3}\frac{dH}{dX}H^2Y^2
    +\frac{1}{2}\frac{d^2\gamma_5}{dX^2}\Big(\tfrac{1}{3}Y^3-\tfrac{1}{2}HY^2\Big)-\frac{1}{4}\frac{d\gamma_5}{dX}\frac{dH}{dX}Y^2,
    \end{multline}
where,
\begin{equation}
\frac{d^2\phi_1}{dX^2}=\frac{d^4\big[H^{-3}]}{dX^4}H^3+ 6\frac{d^3\big[H^{-3}\big]}{dX^3}\frac{dH}{dX} H^2+\frac{d^2\big[H^{-3}\big]}{dX^2}\Bigg(6\bigg[\frac{dH}{dX}\bigg]^2 H+ 3\frac{d^2H}{dX^2}H^2\Bigg)  ,
\end{equation}
\begin{equation}
\frac{d^2\phi_2}{dX^2}=\frac{d^4\big[h^{-2}\big]}{dX^4}H^2+ 4\frac{d^3\big[H^{-2}\big]}{dX^3}\frac{dH}{dX}H +2\frac{d^2\big[H^{-2}\big]}{dX^2}\Bigg(\bigg[\frac{dH}{dX}\bigg]^2+\frac{d^2H}{dX^2}H\Bigg),
\end{equation}
\begin{equation}
\frac{d^2\phi_3}{dX^2}= \frac{d^3\gamma_3}{dX^3}H + 2\frac{d^2\gamma_3}{dX^2} \frac{dH}{dX} +  \frac{d^2H}{dX^2}\frac{d\gamma_3}{dX},
\end{equation}
and,
\begin{equation}
    \frac{d\gamma_5}{dX} = \frac{3}{14}\frac{d^4 \big[H^{-3}\big]}{d X^4}H^4-\frac{3}{5}\frac{d^4\big[H^{-2}]}{dX^4}H^3 + \frac{3}{10}\frac{d\gamma_3}{dx}H^2.
\end{equation}

\section{The Backward Facing Step}\label{app_BFS}
In the limit $\lambda \to \infty$, the logistic step \cref{logistic_eq} resembles the classical backward facing step (BFS) centered at $L/2$,
\begin{equation}h(x) = \begin{cases}H_{\text{in}} & \hspace{2em} 0\le x \le L/2\\ H_{\text{out}} &\hspace{2em} L/2< x \le L \end{cases}.\end{equation}  
Due to the jump discontinuity in $h(x)$ at $x=L/2$, the ELT and $\varepsilon^k$-PLT solutions are not well defined. Since the height gradient is piecewise zero, the solution to the Reynolds equation has pressure contours that are purely one dimensional, and the velocity streamlines are everywhere parallel to $y=0$ instead of navigating around the sharp corners of the step. In contrast, the Stokes solution is able to capture the discontinuity in the height, leading to a similar flow profile as in the logistic step at large $\lambda$.

\begin{figure}
    \centering
     \subfloat[Pressure contours $p(x,y)$]{\includegraphics[width=.95\textwidth]{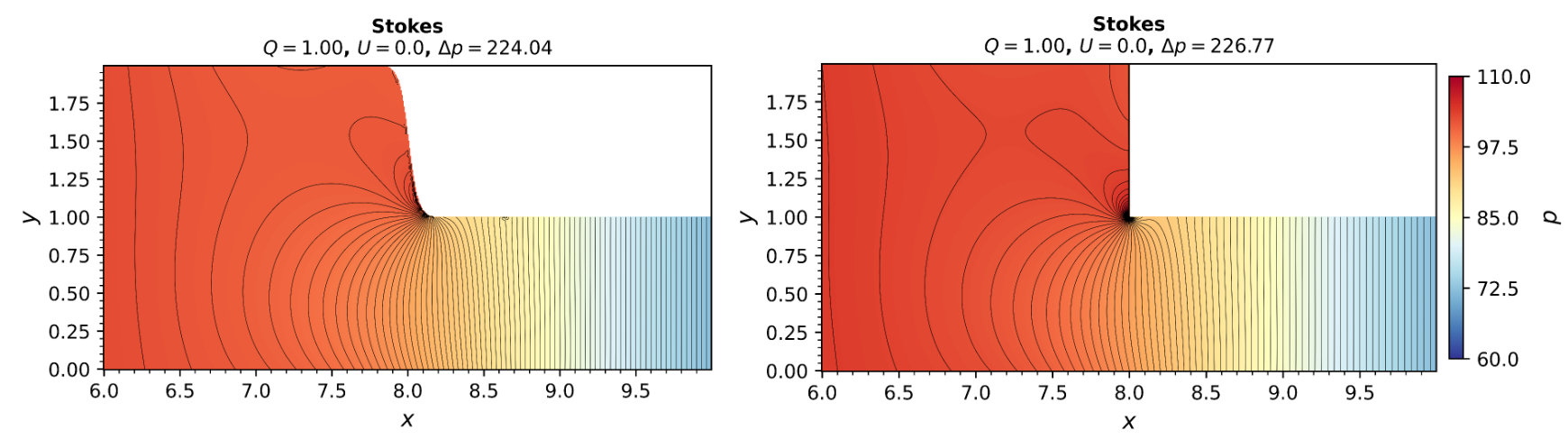}}

    \subfloat[Velocity streamlines $(u,v)$]{\includegraphics[width=.95\textwidth]{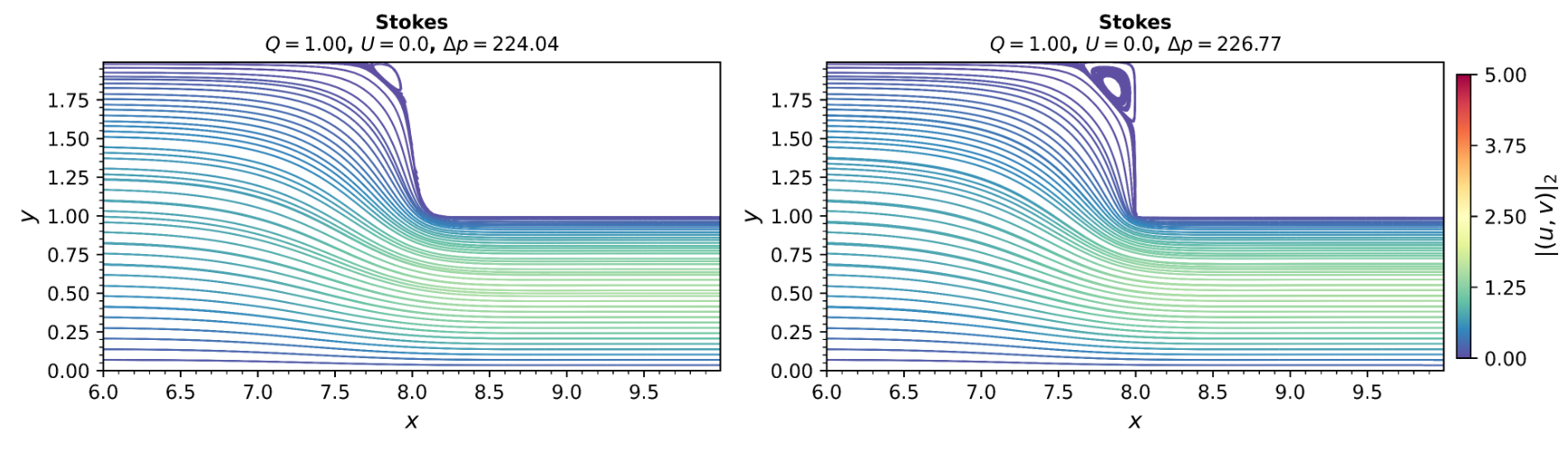}}
   \caption{The Stokes solutions to the logistic step with $\lambda = 32$ (left) and the BFS (right). The logistic step at large $\lambda$ exhibits similar patterns of corner flow recirculation and $\frac{\partial p}{\partial y}$ as in the BFS.}\label{logistic_BFS}
\end{figure}

The Stokes solutions of pressure and velocity for the BFS and the logistic step at $\lambda=32$ are shown in \cref{logistic_BFS}. The two configurations have the same $H_\text{in}/H_\text{out} = 2$, $L=16$, $\nu=1$, $\mathcal{Q} = 1$ and $\mathcal{U}=0$. According to the Stokes solution, the BFS and the logistic step at large $\lambda$ have similar average pressure drops and flow structures. In particular, flow recirculation is significant in the logistic step for $\lambda > 16$, with similar positioning to the recirculation observed in the BFS. Yet, as seen in \cref{logistic_d8}, the Reynolds solution does not exhibit flow recirculation in the logistic step for any $\lambda$, and the VA-ELT and $\varepsilon^{k}$-PLT solutions exhibit flow recirculation at too small of $\lambda$ and with too fast of velocities compared with the Stokes solution.

\bibliography{ReynoldsStokes_bib}
\bibliographystyle{siamplain}
\end{document}